\documentclass[pdflatex,sn-mathphys-num]{sn-jnl}

\usepackage{graphicx}%
\usepackage{multirow}%
\usepackage{amsmath,amssymb,amsfonts}%
\usepackage{amsthm}%
\usepackage{mathrsfs}%
\usepackage[title]{appendix}%
\usepackage{xcolor}%
\usepackage{textcomp}%
\usepackage{manyfoot}%
\usepackage{booktabs}%
\usepackage{algorithm}%
\usepackage{algorithmicx}%
\usepackage{algpseudocode}%
\usepackage{listings}%
\usepackage{lineno}

\theoremstyle{thmstyleone}%
\theoremstyle{thmstyletwo}%

\theoremstyle{thmstylethree}%

\begin{document}

\title[Flat Cell Imaging]{Flat Cell Imaging}


\author*[1,2]{\fnm{Vahid} \sur{Nasirimarekani}}\email{vahid.nasirimarekani@ds.mpg.de}

\author[3]{\fnm{Zuzana} \sur{Ditte}}
\author[1,2,4]{\fnm{Eberhard} \sur{Bodenschatz}}

\affil*[1]{\orgdiv{Max Planck Institute for Dynamics and Self-Organization,}
\normalsize{Am Fassberg 17, 37077, Göttingen, Germany}}

\affil[2]{\orgdiv{Laboratory of Fluid Physics and Biocomplexity},
\normalsize{Am Fassberg 17, 37077, Göttingen, Germany}}
\affil[3]{\orgdiv{Max Planck Institute for Multidisciplinary Sciences},
\normalsize{Am Fassberg 17, 37077, Göttingen, Germany}}
\affil[4]{\orgdiv{Laboratory of Atomic and Solid State Physics, Cornell University, Ithaca, New York},
\normalsize{AIthaca, New York, 14853-2501, USA}}


\abstract{Recent advances in optical technology have significantly enhanced the resolution of imaging of living cells, achieving nanometer-scale precision. However, the crowded three-dimensional environment within cells presents a challenge for measuring the spatio-temporal dynamics of cellular components. One solution to this issue is expansion microscopy, which cannot be used for living cells. Here, we present a method for flattening live cells to a thickness of down to 200 nanometers by confining them between two surface-treated transparent plates. The anti-fouling coating on the surfaces restricts the cells to a quasi-two-dimensional space by exerting osmotic control and preventing surface adhesion. This technique increases the distance between cellular components, thereby enabling high-resolution imaging of their spatio-temporal dynamics. The viability and phenotype of various cell types are demonstrated to be unaltered upon release from flat-cell confinement. The flat cell imaging method is a robust and straightforward technique, making it a practical choice for optical microscopy.}




\maketitle
\newpage
Research in life sciences and many diagnostic methods in medicine primarily rely on imaging of living cells and the interaction of the associated molecules or particles with cellular components. As a matter of size and translucency of living cells, high resolution microscopy of the living cells is a challenging task. Fluorescence staining addresses translucency and enables the acquisition of fluorescence images within a cell or on the cell membrane, which extends to the single-molecule resolution. However, due to the constant spatial relocation of the living cells and their three-dimensional nature, the acquisition of high-resolution images of the cells requires very sophisticated microscopy tools and image processing methods. The best examples of these methods are confocal, multiphoton, total internal reflection, fluorescence resonance energy transfer, lifetime imaging and super-resolution microscopy~\cite{herman2020fluorescence}. In addition, the introduction of expansion microscopy has physically magnified biological samples, enabling the visualization of cellular details~\cite{chen2015expansion,wassie2019expansion}. The approach of expansion microscopy has its drawbacks, such as the need for careful sample preparation, the control of the swelling factor and, above all, the fact that it cannot be applied to living cells. Although some of the techniques developed overcome the obstacle of the 3rd dimension and take images from different planes within a living cell or tissue, the spatial displacement of the living cell and cellular components cannot be avoided. Therefore, one might simply ask whether it is possible to obtain a quasi-two-dimensional configuration of a living cell in order to obtain an image of the cell in which the cell and its cellular components are largely stationary. If this is the case, the entire cell can be visualized simultaneously in a single focal plane using simple widefield microscopy. Logically, this would be possible if a micrometer-sized cell is flattened to a depth of ten or a hundred nanometers, which limits the motility of the cell.
\\
Several methods for flattening living cells based on the application of mechanical forces have been reported. Cell confinement in a chamber with a deformable membrane~\cite{westendorf2010live}, a microfluidics chamber with micro-pillars~\cite{le2014methods}, and the use of an agar overlay~\cite{bretschneider2004dynamic} have been reported.
For example, in the microfluidic method, the height of the pillars defines the maximum depth of the cell between two flat surfaces. The height of the pillars is limited to microfabrication methods, 
hence, the cells were flattened to only a few micrometers at best using the methods described. Apart from the low degree of flattening, flattening living cells by mechanical means is associated with sophisticated setups that require careful preparation and operation. Furthermore, most cells, especially adherent cells, strongly adhere to surfaces. Therefore, two aspects had to be addressed: (i) approaching a quasi-2D flattening by a simple setup and application method (ii) eliminating cell-surface adhesion. Additionally, the living state of the cell is ideally maintained, and the flattened cells can then be re-cultured and reused, an aspect that can barely be achieved with current techniques. 
\begin{figure}[!]
    \centering
    \includegraphics[width=\textwidth]{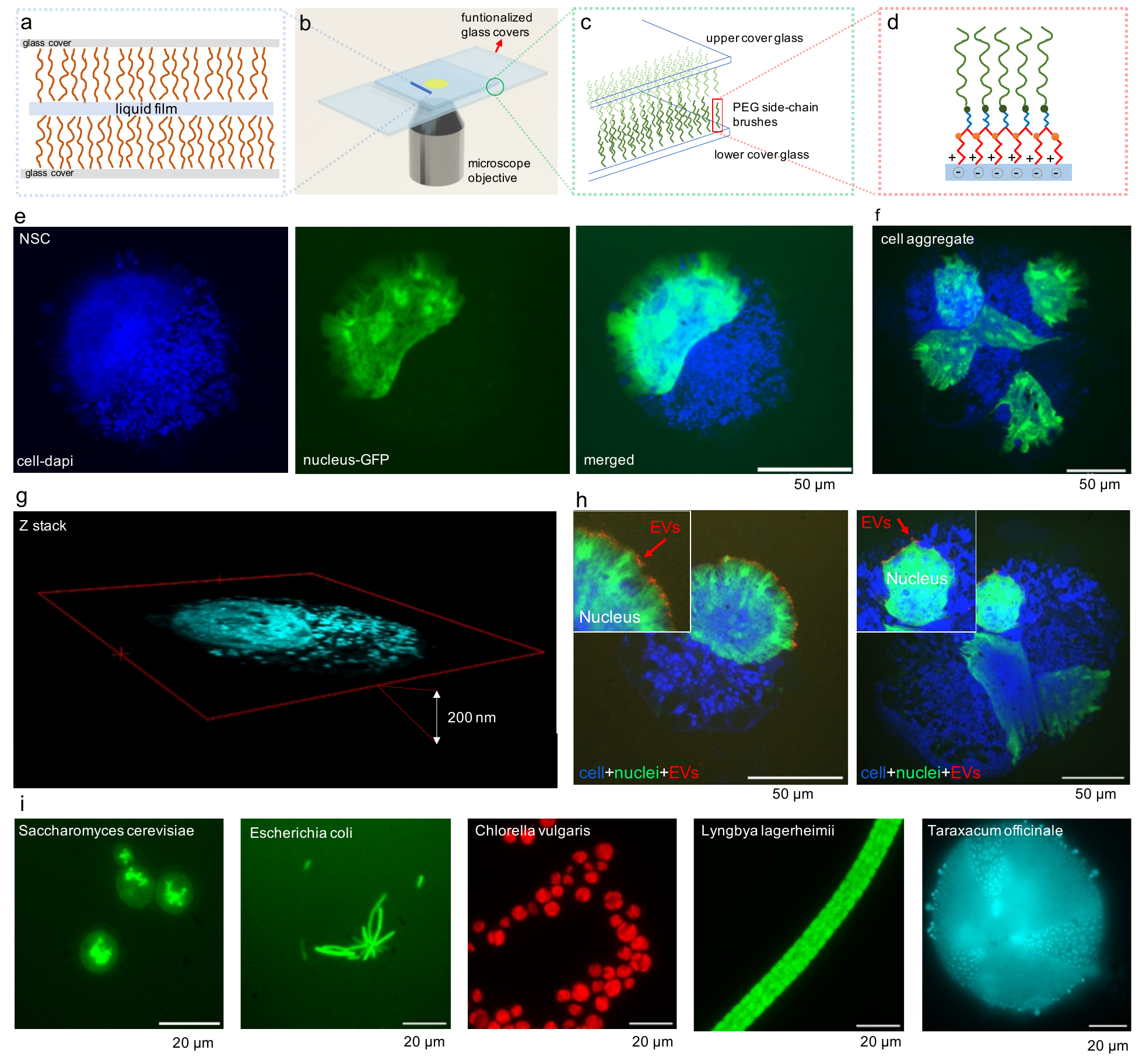}
    \caption{The principal of flat cell imaging and example images acquired by flat cell imaging. a) 2D schematic of the flattening logic of flat cell imaging method, which illustrates the formation of a thin liquid film due to the osmotic and entropic forces of opposing PLL-g-PEG brushes. b-d) 3D schematic of the two opposing surfaces with coated PLL-g-PEG side chains along the chemical structure of the brushes, which are covalently bound to the plasma-activated glass cover. e) An example of a flattened neural stem cell (NSC) with double staining of the cell components and the cell nucleus. f) An example of a flattened aggregation of NSCs, aggregate of 5 cells. g) Z-stack of a flattened NSC showing a depth of 200 nm. h) Example images for using flat cell imaging to observe extracellular vesicle (EV) interaction with the cell nucleus in NSC. i) A series of examples of various flattened cells, namely: Saccharomyces cerevisiae, Escherichia coli, Chlorella vulgaris, Lyngbya lagerheimii and Taraxacum officinale, from left to right.}
    \label{fig1}
\end{figure}
\\
\noindent
We have developed the "Flat Cell Imaging" method which enables us to flatten living cells down to 200 $nm$, depending on the cell type, whereby a 2D expansion of the living cells by up to 9 times is achieved.
This was simply accomplished by flattening living cells in between two high-precision glass microscopy slides functionalized with a graft copolymer of poly(L-lysine) backbone and poly(ethylene glycol) side chains (PLL-g-PEG), which served as two flattening surfaces (Figure~\ref{fig1}.a-d). The coating offers two main properties: (i) It blocks any plasma membrane attachment to the surfaces, as it is known to repel proteins and cell membranes, known as an antifouling property~\cite{elbert1998self, csucs2003microcontact, lussi2006pattern, marie2006use}; (ii) two coated surfaces facing each other cause a high ratio spreading of the liquid between them, resulting in a nanometer depth of the flattened cell (Figure~\ref{fig1}.a). 
\\
The Pll-g-PEG coating has a brush-like structure with well characterised protein absorption~\cite{pasche2003poly} and friction coefficient measurements~\cite{perry2009tribological}. In our method, the living cell is flattened between two opposing brush-like layers that have a high degree of lubricity. Lubricity is provided by a combination of osmotic pressure and interfacial penetration resistance within the facing brushes, known as entropic forces. Osmotic pressure provides a significant reduction in dissipative forces, resulting in an ultralubric interface~\cite{benetti2019using,abdelbar2023polymer}. This unique property of the coated surface facing each other allows the flat cell imaging method to flatten the cells and approach a quasi-2D shape without the need to apply external mechanical forces (Extended Data Fig.1). 
\\
The flat cell imaging method was tested on various cell types, ranging from mammalian cells to amoebae and prokaryotic cells. The list of the cells and the flattening ratios ($f$, as a measure of the 2D shape in relation to its original shape, Extended Data Fig.2) are summarised in Table~\ref{tab1}. Cells with exposed plasma membrane (without cell wall or extracellular matrix covering the cell), such as mammalian cells, have high degrees of flattening. As an example, in the case of neural stem cells (NSC), which show a round shape around 10~$\mu m$ of diameter in a suspension, the single cells or cell aggregates were flattened up to $90~\mu m$, with a polar positioning of the cell nucleus (Figure \ref{fig1}.e,f, the nucleus is shown in green color). Z stack imaging of a flattened NSC shows 200~$nm$ depth of the cell (Figure \ref{fig1}.g). As the cell depth is comparable to the size of the extracellular vesicles (EVs, around 150~$nm$), the method was used to localize the interaction of the EVs with NSCs (Figure \ref{fig1}.h, EVs are seen to be localized at the periphery of the cell nucleolus). This demonstrated that flat cell imaging visualizes the interaction of nanometer-sized particles with cellular components imaged in a single plane.    
\begin{table}[h!]
\centering
\caption{List and flattening ratio of cell lines and organisms subjected to the flat cell imaging method}\label{tab1}%
\begin{tabular}{@{}llll@{}}
\toprule
cell or organism type & $f(min-max)$ & fluorescent & cell origin \\
\midrule
NSC    & 6 - 9  & cell+nucleus & mouse \\
HL-1    & 6 - 9  & cell+nucleus & mouse \\
Panc-1    & 6 - 9  & cell+nucleus & human  \\
NIH 3T3    & 6 - 9  & cell+nucleus & mouse 
\\
MDCK    & 6 - 9  &cell+nucleus  & canine  \\
HeLa   & 6 - 9  & cell+nucleus & human  \\
Dictyostelium discoideum& 6 - 9  & Actin-GFP  & amoeba  \\
Taraxacum officinale   & 4 - 6  & auto fluorescence & plant  \\
Lamium purpureum    & 4 - 6  & auto fluorescence & plant  \\
Euphorbia myrsinites    & 4 - 6  & auto fluorescence & plant  \\
Saccharomyces cerevisiae  & 3 - 4  & auto fluorescence & yeast  \\
Chlorella vulgaris  & 2 - 3  & auto fluorescence & algae  \\
Lyngbya lagerheimii  & 2 - 3  & auto fluorescence & cyanobacteria  \\
Escherichia coli    & 2 - 3 & auto fluorescence & bacteria  \\

\botrule
\end{tabular}
\end{table}
\\
The flattening ratio $f$ decreased in plant cells, which are stiffer due to the outer cell wall in addition to the plasma membrane (Extended Data Fig.3). Another control parameter that affects the flattening ratio is the initial size of the cell. As the initial size of the cell decreases, the flattening ratio also decreases. In smaller cells such as E.coli, Lyngbya lagerheimii and Saccharomyces cerevisiae, $f$ value decreased, showing maximum 4 times flattening (Table~\ref{tab1}, Figure \ref{fig1}.i). From this, it can be deduced that the degree of flattening of the cells depends both on the rigidity of the cell and its intrinsic size. 
Nevertheless, the rigidity of a living cell changes depending on the developmental stage of the cell~\cite{starodubtseva2011mechanical}, which explains the range of flattening ratios measured in the experiments. 
\\
Although flat cell imaging is primarily aimed to provide 2D imaging of cells, it is of great advantage to maintain the live state of the cells so that they can be collected and cultured for further studies or microscopy steps. In this context, we examined the flattened cells after flat cell imaging (Figure \ref{fig:fig2}.a). The collected cells underwent a recovery phase in which they retained their original shape and size (Figure \ref{fig:fig2}.b, Extended Data Fig.4). The collected cells were then cultured again under the required culture conditions and the cell cycle, differentiation, proliferation and viability of the collected cells were examined in the case of four different cell lines (see Methods). 
\\
The life cycle of the \textit{Dictyostelium discodium} (D.d), a social amoeba cell that undergoes cell aggregation through chemical signaling, differentiates and forms spores under starvation conditions~\cite{kessin2001dictyostelium}, was studied. The collected flattened \textit{D.d} cells were subjected to starvation, and the results showed the normal cell cycle completion (Figure~\ref{fig:fig2}.c,d, Extended Data Fig.5, the formation of slugs and fruiting bodies are emphasized, which involves cell differentiation into spore and stalk cells~\cite{kessin2001dictyostelium}). In addition,
differentiation of NSCs, activated by EVs produced by choroid plexus cell line~\cite{ditte2022extracellular}, was tested after recovery of collected flattened cells. Incubation for 24 hours showed that differentiation was not affected by 2D-flattening and that the cells were differentiated into neurons and astrocytes (Figure \ref{fig:fig2}.e). The presence of astrocytes was confirmed by immunofluorescence staining for the astrocyte marker (Extended Data Fig.6). In addition, cell proliferation of fibroblasts and pancreatic cancer cells showed regular cell proliferation rates (Figure \ref{fig:fig2}.e, Extended Data Fig.7). 
\\ 
Flattened cells showed viability values of 85\% to 96\% at flattening times below 15 minutes, with lower values for NSC and higher values for \textit{D.d} cells. (Figure \ref{fig:fig2}.f). However, the viability values dropped to about 60\% at a 30 mins flattening time. This suggest that shorter flattening time would be beneficial in the case of desiring to re-culture them flattened cells. Viability is affected by mechanical stress on the cell (osmotic pressure), the lack of gas exchange through the glass slides, evaporation from the edges of the glass slides, and performing experiments at room temperature. We hypothesize that if an imaging time of more than 15 min is desired, then viability can be improved by optimizing the flattening ratio and gas permeability of the flattening surface.
\begin{figure}[!]
    \centering
    \includegraphics[width=\textwidth]{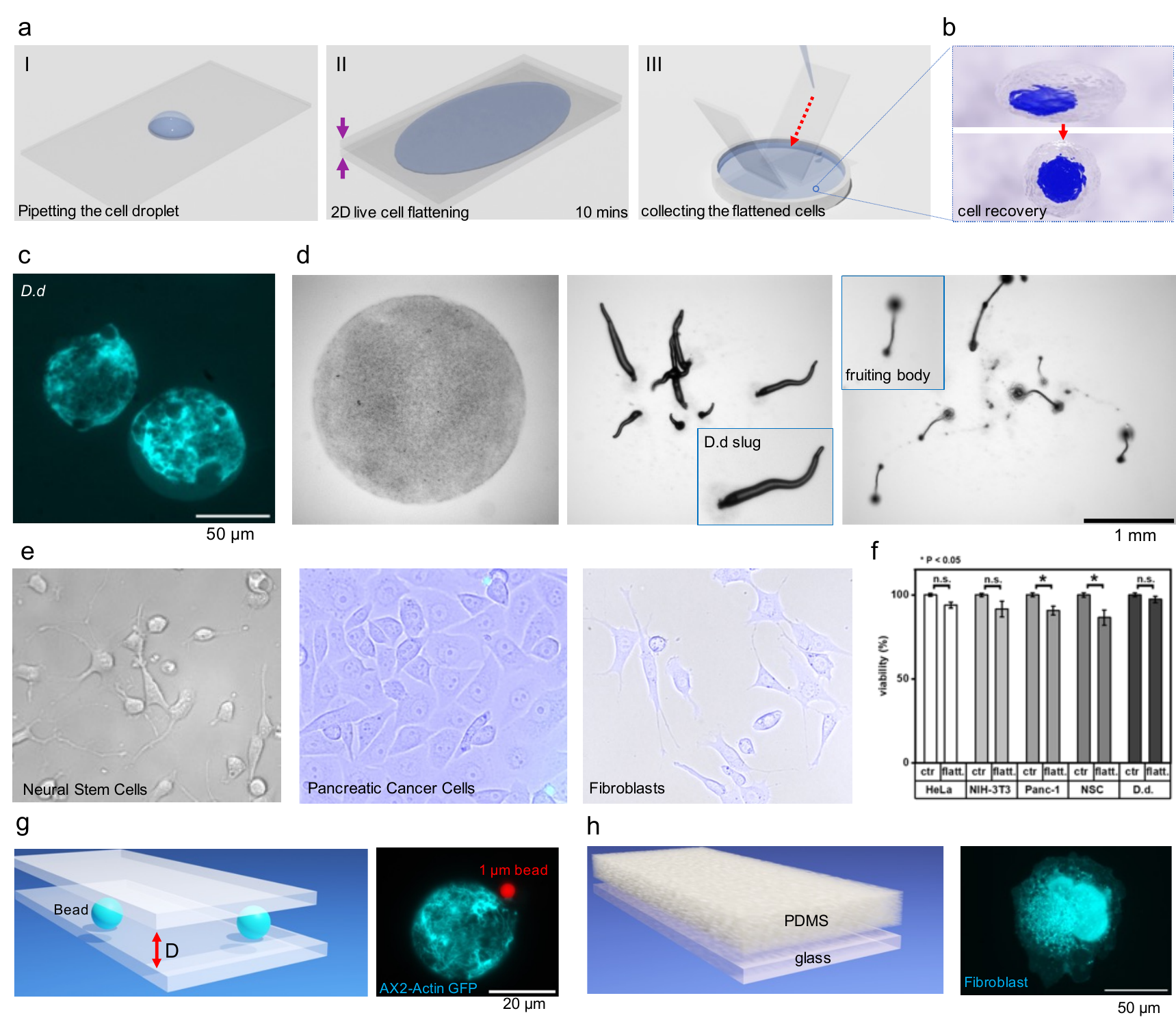}
    \caption{Recovery of flattened cells, measurement of their life cycle and viability. a) Schematic representation of the steps for flattening the cells and collecting the flattened cells. b) Schematic representation of cell recovery of the collected flattened cell. c) Examples of the flattened \textit{D.d} cells with AX2-Actin-GFP in cyan color. d) The life cycle of re-cultured \textit{D.d} cells after being flattened for 15 mins, the slug and fruiting body formations are highlighted. e) Examples of three different cell lines taken after 15 minutes of flattening: differentiated neural stem cells (NSCs) through extracellular vesicles (EVs), proliferating pancreatic cells, and fibroblasts, from left to right. f) Viability values of the flattened cells measured after 24 hours of cells that were flattened for 15 minutes. g) The addition of beads as spacers controls the flattening ratio. The beads control the distance between two glasses. h) Demonstration of the use of porous and gas-permeable polydimethylsiloxane (PDMS) as an upper flatening plate.}
    \label{fig:fig2}
\end{figure}
\\
We have developed a control to reduce the degree of flattening by adding nano- or micro-sized monodisperse particles to the cell suspension that act as spacers between the flattening surfaces (Figure \ref{fig:fig2}.g). 
In eukroyatic cells with exposed plasma membrane, the reduction of $f$ is beneficial as the achieved flattening ratios are considerably high (up to $f=9$, Table.1) and can be reduced if necessary. 
In addition, we have questioned the possibility of using a porous polymeric material for the upper flattening surface, instead of glass covers. Polydimethylsiloxane (PDMS) was tested due to its biocompatibility and possibility to apply Pll-g-PEG coating~\cite{marie2006use}. Moreover, PDMS is an elastic material which can be used in different thicknesses, stiffness and porosity~\cite{lamberti2014pdms}. We have observed that PDMS layer give similar results as reported in Figure~\ref{fig1}.e,f and Table~\ref{tab1}. Therefore, we propose that PDMS can be used as a gas-permeable membrane, which would be beneficial for the viability  of the flattened cells and longer imaging times.
\\
In this article, we have introduced the flat cell imaging method, a quasi-2D flattening method that enables high-resolution imaging of living cells. The flat-cell imaging method uses a simple and novel approach that takes advantage of the antifouling and lubricating property of the brush-like structure of the Pll-g-PEG coating to achieve non-invasive 2D flattening of living cells. By demonstrating the applicability of the flat cell imaging method for various cell types or organisms, we aimed to present reference data to communicate the potential of the method for live cell imaging. Furthermore, we showed that the phenotype and cell cycle of flattened cells are unchanged. Together with high viability rates, flat-cell imaging offers the possibility of multiple imaging steps from the same cell population. We envision this method to be used for monitoring the interaction of the particles or molecules with the cell membrane or cellular compartments in real time.  Furthermore, we foresee that the mechanical stress caused by the flattening of the cells could lead to a disruption of the cell membrane, suggesting an applicability for intracellular delivery or extraction.
\bibliography{sn-bibliography}
\section*{Methods}
\subsection*{Preparation of Pll-g-PEG coated glass slides, and  PDMS layer}
Microscope coverslips (purschased form Paul Marienfeld GmbH \& Co. KG 
, 24*60 mm with thickness of 0.170 mm ± 0.005 mm and weight of 0.605 ± 0.001 gr) were cleaned by washing with 100\% ethanol for 10~min and rinsing in de-ionized water afterwards. They were further sonicated in acetone for 30~min and followed by incubation in 96\% ethanol for 10~min. Last step was 2 hours of incubation in a 2\% Hellmanex III solution (Hellma Analytics). Finally, cover slips were extensive washed in de-ionized water and dried with a dry and filtered airflow. All cleaning procedures were performed at room temperature. The cleaned cover slips were stored in 4$^{\circ}$C fridge. Prior to application of the Poly-L-lysine-graft-polyethylene glycol (PLL-g-PEG) coating, the celaned cover slips were activated in oxygen plasma (FEMTO, Diener Electronics, Germany) for 3~min at 0.5 mbar. The coating solution with final concentration 0.1 mg/ml of the PLL-g-PEG (SuSoS AG, Switzerland) in 10 mM HEPES (pH 7.4, at room temperature) was prepared. PLL-g-PEG stock solution (1mg/ml) in 10 mM HEPES was filtered through 0.2 $\mu m$ filter and stored at 4$^{\circ}$C. The coating solution was applied as drops on a clean surface (parafilm) and the activated glass slides were put on it and incubated for 30~min in a ventilated hood. Remaining liquid was removed and slides were dried by the air flow. The coated slides were used right away or stored in 4$^{\circ}$C for later uses. 
\subsection*{Preparation of PDMS layer}
 Polydimethylsiloxan (PDMS) was crosslinked by mixing the PDMS (SYLGARD 184, purchased from Sigma Aldrich) with crosslinker (SYLGARD 184, silicon elastomoer curing agent), in 1 to 10 weight ratio of curing agent to PDMS. The well-mixed mixture was poured onto a coverslip in a petri dish. The glass was used to achieve a surface flatness similar to that of the previously used cover glasses. The petri dish, partially filled with the PDMS mixture, was first degassed in a desiccator and then incubated for one hour in an oven preheated to 75$^{\circ}$C. The cross-linked PDMS was peeled off the glass slide and cut into a rectangular shape. All steps were carried out under a laminar flow hood to avoid dust accumulation on the PDMS surfaces. Subsequently, the Pll-g-PEG coating was applied according to the procedure described in the previous section. 
\subsubsection*{Cell Culture}
\textbf{Plant cells:} Individual plant cells (Taraxacum officinale, Lamium purpureum and Euphorbia myrsinites) were isolated using a mechanical method~\cite{gnanam1969photosynthetic}. In short, the plants were cleaned and scraped with a scalpel and then ground manually in an grinding medium (20 $\mu M$ sucrose, 10 $\mu M$ MgCl2, 20 $\mu M$ Tris-HCl buffer, pH 7.8), followed by sequential filtration thought the cell strainer (100 $\mu m$, 70 $\mu m$ and 37 $\mu m$, all BD Falcon). Cells were three times washed in grinding medium by gentle centrifugation (5 min, 200g) and were collected in an Eppendorf tube. 
\\
\\
\textbf{Mamalian cells:}
\\
All mammalian cell lines were cultured and incubated in a humidified incubator with 95\% air, 5\% CO2 at temperature of  37$^{\circ}$C. 
\\
\textit{mouse neural stem cells (NSCs):} NSCs isolated from subventricular zone~\cite{walker2014one} were grown in Neurobasal Medium supplement with 2\% B27, 1\% N2 supplement, 1x GlutaMAX, 50 units/ml Penicillin/Streptomycin (all Gibco), 20ng/ml purified mouse receptor-grade epidermal growth factor (EGF), and 20ng/ml recombinant bovine fibroblast growth factor (FGF-2) (both purchased from Peprotech). 
\\
\textit{Mouse embryonic fibroblast cells (NIH 3T3), human pancreatic cancer cells (Panc-1) human epithelial cancer cells (HeLa)}: all the three cell lines were grown in DMEM (Dulbecco's Modified Eagle Medium) supplemented with 10\% FBS, 1x GlutaMAX and 50 units/Penicillin/Streptomycin (all Gibco™, purchased from ThermoFisher). 
\\
\textit{Mouse cardiomyocyte (HL-1)}: The cells were grown in Claycom medium, supplement with 10\% FBS, 1x GlutaMAX, 0.1$\mu M$ Norepinephrine and 50 units/Penicillin/Streptomycin (all Sigma). Madin-Darby canine kidney cells (MDCK) were grown in MEM supplemented with 10\% FBS, 1x GlutaMAX and 50 units/Penicillin/Streptomycin (all Gibco™, purchased from ThermoFisher)
\\
\\
\textbf{Social amoeba-\textit{Dictyostelium discoideum (D.d)}:}
\\
The life cycle experiments were performed with D.d AX2-214 cells, kindly provided by Prof. Günther Gerisch (MPI for Bio-chemistry, Martinsried, Germany). Cells were grown in HL-5 medium (35.5g of Formedium powder from Formedium Ltd, England, per liter of double-distilled water, autoclaved and filtered) at 22$^{\circ}$C on polystyrene Petri dishes (TC Dish 100, Sarsted, Germany) and harvested when they became confluent. Before the experiments, the cells were centrifuged and washed two times with phosphate buffer (2g of $KH_{2}PO_{4}$ and 0.36g of $Na_{2}HPO_{4}.H_{2}O$ per liter at pH 6.0, autoclaved, both from Merck, Germany). The centrifuged cells were resuspended in 10 ml of the same buffer and were collected in a final solution of 200 $\mu$l with a cell count of 1 million per micro-litter. 
\\
The AX2-Actin-GFP cells were provided kindly by Prof. Günther Gerisch (MPI for Bio-chemistry, Martinsried, Germany) as well. The cells are genetically modified with actin-GFP plasmid and selection marker of G418 resistance gene (Geneticin) was used as 0.1 mg/ml in the culture for selection purpose.
\\
\\
\textbf{Bacteria:} 
\\
\textit{Cyanobacteria, Lyngbya lagerheimii:} the specie was obtained from The Culture Collection of Algae at Gottingen University and seeded in T175 culture flask with standard BG-11 (Gibco™, purchased from ThermoFisher) nutrition solution. The culture medium was exchanged for fresh BG-11 every four weeks. Cultures were kept in an incubator with an automated 12 h day (30 \% light intensity (around 20 $\mu$E), 18$^{\circ}$C) and 12 h night (0 \% light intensity, 14 $^{\circ}$C) cycle, with a continuous 2h transition.
\\
\textit{Escherichia coli:} 25 ml of LB medium (Invitrogen, 12780052) was pipietted into a 125 ml Erlenmeyer flask and 25 $\mu$l ampicillin (stock solution: 50 mg/ml, Carl Roth, K0291) was added to the LB medium. The medium was inoculated with NEB 5-alpha Electrocompetent E. coli (NEB, C2989K) and incubated overnight in a shaking incubator (100 rpm) at 37° C.
\\
\\
\textbf{Yeast:}
\\
\textit{Saccharomyces cerevisiae:} was purchased as commercial yeast from supermarket and dissolved in PBS (Phosphate-buffered saline, ph 7.4) and supplemented with Glucose. It was further diluted in PBS to achieve desired cell count.
\\
\\
\textbf{Algae:}
\\
\textit{Chlorella vulgaris:} specie was obtained form X-lab in Göttingen, kindly provided by Dr. Dirk Gries. It was cultivated in aerated vessel with applying commercial plant fertilizer.
\subsubsection*{Live Cell Staining}
Prior the staining procedure each type of the cells were passaged to obtain single cell suspension. Moreover, the cell concentration was determined using the hemocytometer (Nexcelon, Bioscience). 
All dyes were used according to the manufacture's guidelines. Cell pellet ($2*10^5$ cells/ml) were re-suspended in pre-warmed PBS with cell Trace Violet staining solution (5$\mu$M, Thermo Fisher Scientific) and incubated for 20min, 37$^{\circ}$C. First, cells were spun down (5min at 200×g) and the formed pellet was resuspended in fresh cell culture corresponding media. Then, they were incubated for 5 min, 37$^{\circ}$C to absorb any unbound dye. After another centrifugation step (5min, 200g), the cell pelet was resuspended in corresponding media with NucSpot® Live 488 (final concentration 1x) (Biotium) and incubated for 10min, 37$^{\circ}$C.
\subsection*{EVs purification and labeling}
\textit{Purification:} EVs  were purified by differential ultra-centrifugation from the cell conditional media of the immortalized Z310 rat choroidal epithelial cells. Briefly, after 48 hours of cultivation of the cells in serum free medium, medium was collected and centrifuged in a stepwise order at 300g for 10min,  at 2k g for 20 min (Eppendorf 5702R) and 10k g for 40 min (Eppendorf 5417R). the formed supernatant was concentrated by Vivaspin (300 000 MWCO, Sartorius, Göttingen, Germany). Final supernatant was subjected to the ultra-centrifugation at 100k g, for 60min, 4$^{\circ}$C (Sorvall WX-Ultra 80, Thermo Fisher Scientific, USA) in a TH-660 rotor (Thermo Fisher Scientific). The resulting pellet was resuspended in ice cold PBS and centrifugation was repeated under the same condition. Pellet containning EVs was resuspended in PBS and stored at -80$^{\circ}$C.
\\
\textit{Labeling:} EVs were fluorescently labelled by specific primary antibody against annexin 2 (Abcam, Ab178677) overnight at 4$^{\circ}$C, then incubated with corresponding secondary antibody Alexa Fluor 555 conjugated donkey anti-rabbit IgG (Life Technologies, A31572), for 1h at the room temperature. Moreover, the samples were washed three times with PBS (0.1\% BSA, 0.1\% Triton), (Amicon ultra, UFC510096) to remove unbound dye.

\subsection*{Flat Cell Imaging}
Before imaging, to prevent potential biological contamination, the PLL-g-PEG glasses underwent a 30 min UV light treatment within a cell culture hood. A droplet of media with suspended cells (around 4 $\mu l$) was pipetted on the surface of the functionalized glass cover-slip. 
The second cover slip, with the functionalized side facing the droplet, was manually placed onto the drop. The liquid film in between the glasses forms in less than a second and spreads further totally roughly for 20 seconds. The glass cover-slips contianing the flattened cells was introduced then to the microscope with a climate chamber maintaining temperature, humidity and $CO_{2}$ concentration of the cells. The transfer of the sample to the microscope was done fast to avoid affecting the cell viability. The flattening step was done inside a hood to avoid deposition of dust and contamination on the glass slides. 
\subsection*{Immunofluorescence (IF) Staining}
After 24h, cells were fixed in 4 \% formaldehyde, washed with PBS ( 0.1 Triton, 0.1 \% BSA) and blocked at the room temperature for 30 min (4\% goat serum in 0.1\% Triton in PBS).Then cells were incubated with primary antibody against GFAP (Abcam, ab4674) overnight at 4°C. Three washing steps with PBS (0.1\% BSA, 0.1\% Triton), were followed by incubation with secondary antibody Alexa Fluor 488 conjugated goat anti-chicken IgG, Alexa Fluor 555 conjugated donkey anti-rabbit IgG, for 1h at room temperature. After three washes coverslips were mounted with mounting medium with DAPI (Vectashield, Vectorlabs) and images were acquired using a Leica DMI 6000B fluorescence microscope, with LAS X software (Leica Microsystems, Germany).
\subsection*{Collection and re-culture of the flattened cells}
After desired flattening time, the two glass slides were gently separated by using a carbon steel surgical scalpel (precleaned with ethanol and washed with mQ water). The separated glasses were hold by hand with an angle to permit flowing of the liquid to the bottom of the glass and washing with cell buffer. For smaller volumes, the the buffer was pipetted and collected on the glass-slides and introduced inside an Eppendorf tube. In the case of bigger volumes the cell solution was washed directly inside a Petri dish. The collection procedure was done under sterilised hood at the room temperature. Due to evaporation arisen from the air flow inside the hood the collection step was done as fast as possible.
\\
The collected cells were cultivated according to the standard culture conditions for each cell line (as previously reported). Adherent mammalian cells such as NIH-3T3, HeLa, NSC, Panc-1 and D.d cells were observed for their ability to attach to the surface and typical morphology. 
\subsection*{Life cycle studies of flattend \textit{Dictyostelium discodium (D.d)}}
For life cycle studies of \textit{D.d}, the cells were prepared in the phosphate buffer to initiate the starvation process. The flattening was done for 15 min and the cells were then collected in the same buffer. The collected cells were centrifuged to achieve a cell concentration of 3 million/ml. A drop of the cell suspension was dropped on 1.5\% phosphate agar inside a Petri dish with a lid. The time-lapse phase contrast images of the cells were acquired over 24 h to record the life cycle of the cells (SI-V2). The microscopy was conducted at room temperature. 
\subsection*{Differentiation of NSC by EVs}
NSC cells after being confined for 10 mins were collected and seeded on the cell culture dish in corresponding cell media. Immediately, EVs choroid plexus origin were added to the cell culture media (total EV protein concentration 76 $\mu g/ml$) and mixed by pipetting. As a control, NCS which have not been confined were cultured under the same condition with EVs, to evaluated if confinement have influence to the NSC differentiation capacity. After 24 hours of co-incubation bright field images of live cell were taken. Both NSC, confined and non-confined, attached to the surface and formed organized cellular network in presence of EVs. Prolong cultivation for 48 hours revealed cells with typical shape of astrocytes and neurons as described previously~\cite{ditte2022extracellular}.
\subsection*{Cell Viability}
All cell types were after flattering washed with the corresponding buffer, collected in Eppendorf tube and centrifuged at 300×g for 5 min. Cells were re-suspended in corresponding media and seeded on 96-well flat-bottom  tissue cell culture plate of good optical quality (Ibidi, Germany). The microscope images were taken after 24 and 48 hours respectively. Dimethyl sulfoxide (DMSO) was used to compare how the cells respond to the presence of a high concentration of DMSO and flattening. The cells responded by significant viability reduction to the presence of DMSO, in contrast to flattening, which they tolerated well.
\\
After 48h recovery period cell viability was assessed by MTT cell viability assay. All steps were performed according manufacture’s instructions (Merck, Colorimetric MTT kit for cell survival and proliferation). Absorbance was detected with ELISA plate reader (Infinite M2000 Pro Tecan) with a test (570 nm) and reference (630 nm) wavelength. Obtained data were plotted by OriginPro 2021 (OriginLab, USA), values are presented as mean± SD from three independent experiment ($*p < 0.05, **p < 0.01$, Student's t test). For Trypan Blue assay cells underwent analysis immediately after flattering according manufacture’s instructions (Sigma) and the number of live/dead cells was determined by hemocytometer (Nexcelon, Bioscience).
\subsection*{Adding beads as spacer}
Wide range of commercial beads (Malvern Panalytical and Bangs Laboratories, Inc, with the size range of 300 nm to 5 microns) were used as spacer in between two glass slides. However, the  fluorescent beads were preferred for better visualization of them. The desired beads were diluted in the cell culture medium or buffer corresponding to the cell type. The mixture was shortly vortexed before conducting the Flat Cell microscopy. For micron-sized particles the mixture was few times pipetted inside an Eppendorf tube to achieve a better suspension of the particles inside the cell solution and avoid the clustering of them in the bottom of the Eppendorf tube.  

\subsection*{Microscopy and Image Analysis}
\textbf{Flat Cell Imaging:} Image acquisition was performed using an inverted fluorescence microscope Olympus IX-71 with a 60$\times$-oil objective (Olympus, Japan), with various binning and window size depending on the experiment. For excitation, a Lumen 200 metal arc lamp (Prior Scientific Instruments, U.S.A.) was applied. The images were recorded with a CCD camera (CoolSnap HQ2, Photometrics). The frames for videos  were acquired with at different Hz, for fast speeding acquition 300 ms intervals was used as the minimum possible interval. 
\\
\textbf{Other microscopy imaging:} The images for cell viability and proliferation evaluation, NSC differentiation were acquired using a Leica DMI 6000B fluorescence microscope, with LAS X software (Leica Microsystems, Germany, Fluoreszenz-Imaging-System Leica AF6000).
\\
Olympus CKX41 microscope with a mounted Axiocam MRm camera and a phase contrast ring was used for life cycle studies of \textit{D.d}. 
\\
The microscopy for droplet spreading dynamics for non-treated, cleaned and functionalized glasses was conducted by using the eiss Axio Zoom V.16 microscope equipped with a Hamamatsu ORCA-Flash 4.0 digital camera and a Zeiss PlanNeoFluar1×/0.25  objective. 
\\
\textbf{Image analysis}
ImageJ software was used for the image processing and analysis of the acquired images.
\subsection*{flattening ratio, $f$:}
\textit{Mammalian, amoeba, yeast, plankton and plant cell:} Flattening ratio is a measure of the compression of a spherical cell flattend to a eliposoid or irregular cell shape. The notation, $f$ is measured as $f=D_{2D}/d_{is}$, wherein $d_{2D}$ is the average diameter of the cell in 2D shape (calculated as $f=(f_a+f_b)/2$, $f_a$ and $f_b$ represents the long and short diameter of tge flattened cell). In this connection it is of note that cells are generally spheroids. In the case of yeast cells only the main body of the cell was considered for measurements. 
\\
\textit{E-coli and filamentous cyanobacteria:} in the case of the bacteria $f$ only was calculated as $f=d_{id}/d_{2D}$. As it was not possible the measure the initial length of the bacteria the comparison is only done in diameter. 

\section*{Acknowledgment}
The authors are grateful to Prof. Gregor Eichele for the valuable initial discussions that encouraged the authors to think of a new ways for high resolution microscopy. The authors thanks the Max Planck Society for the financial support.
\section*{Author contributions}
V.N and Z.D conducted the experiments, analysed of the data and wrote the first draft. E.B and V.N conceived the  flattening idea and the microscopy method. All authors discussed the results and contributed to the final manuscript. 
\section*{Conflict of interest}
V.N and E.B hold a patent on the flat cell imaging method (PCT/EP2023/080067). In addition, the same authors hold two separate patents on intracellular delivery and extraction using the cell flattening method presented in the article (PCT/EP2023/080082, PCT/EP2023/080079, respectively). 
The other authors declare no competing interests. 
\setcounter{figure}{0}
\makeatletter 
\renewcommand{\thefigure}{\@arabic\c@figure}
\makeatother
\newpage
\section*{Extended Data}

\subsection*{comparison of droplet spreading on different glasses}
\begin{figure}[h!]
    \centering
    \includegraphics[width=\textwidth]{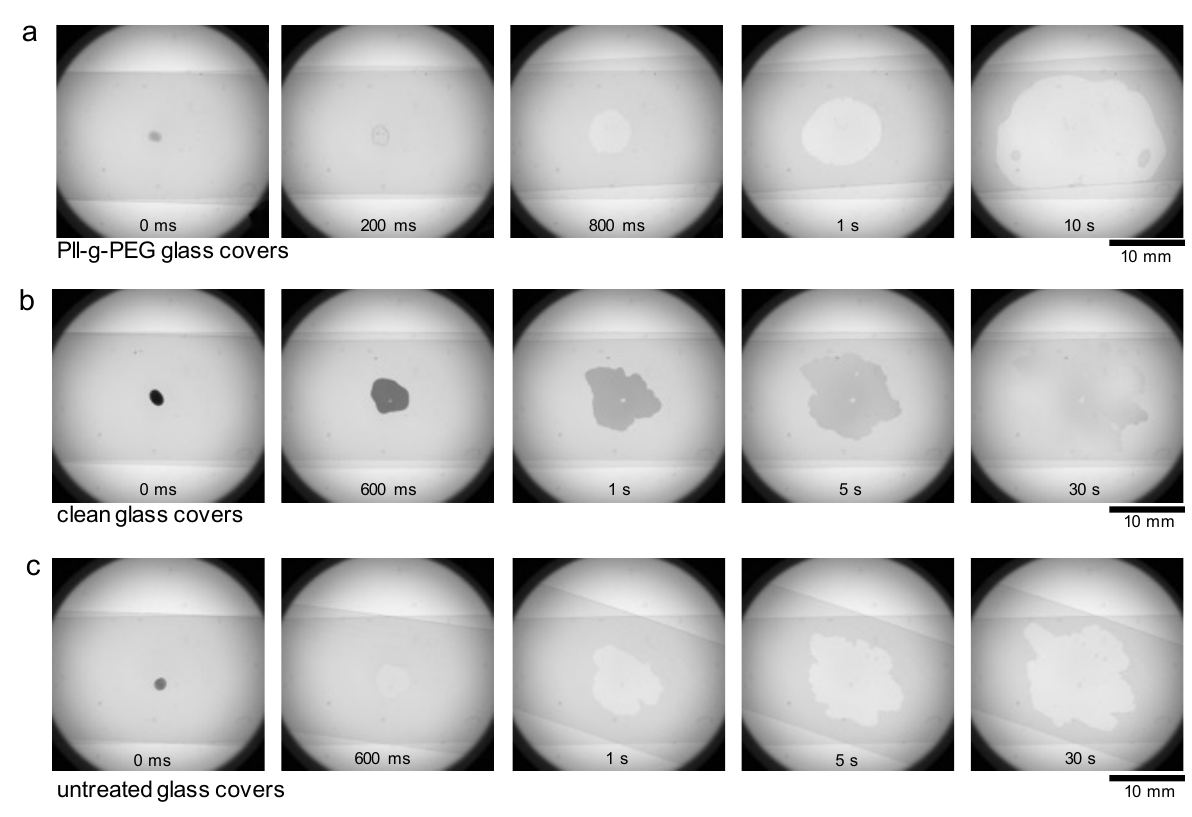}
    \caption{Comparison of liquid film formation between two glass coverslips with three different types of treatment: untreated glass coverslips, cleaned and PLL-g-PEG-coated. a-c) The time-lapse images of the spread for the PLL-g-PEG, cleaned and untreated coverslips. }
    \label{fig:enter-label}
\end{figure}
\newpage
\subsection*{Flattening ratio}
\begin{figure}[h!]
    \centering
    \includegraphics[width=0.8\textwidth]{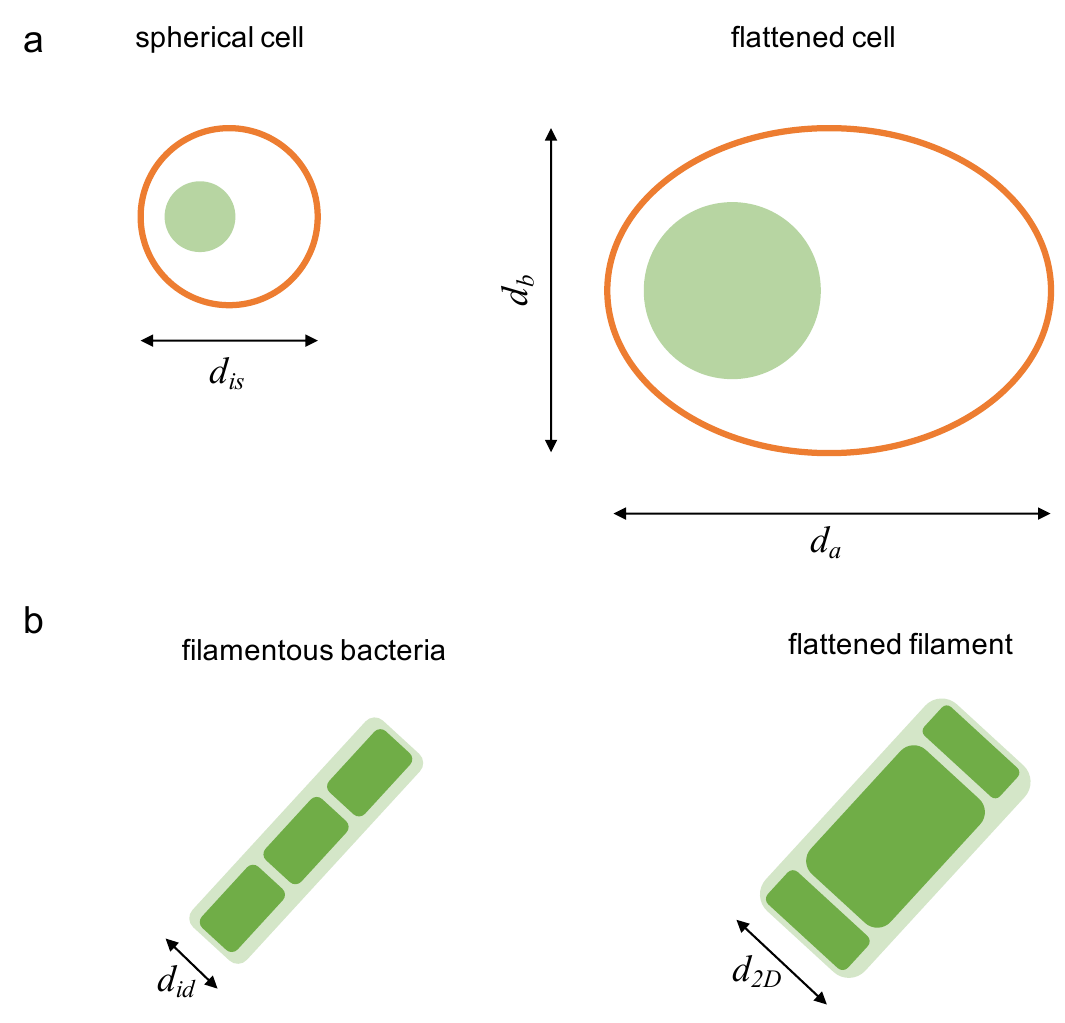}
    \caption{Flattening ratio calculations for flattened cells. a) schematics for flattening ratio measurements of mammalian and amoeba cells, all considered as sphere cells. b) schematics for e-coli and filamentous cynobacteria, the flattening ratio is only considered for the diameter of the bacteria and the length of it is not measured.  }
    \label{fig:enter-label}
\end{figure}
\newpage
\subsection*{Comparison of non-flattened and flattened plant cells}
\begin{figure}[h!]
    \centering
    \includegraphics[width=0.8\textwidth]{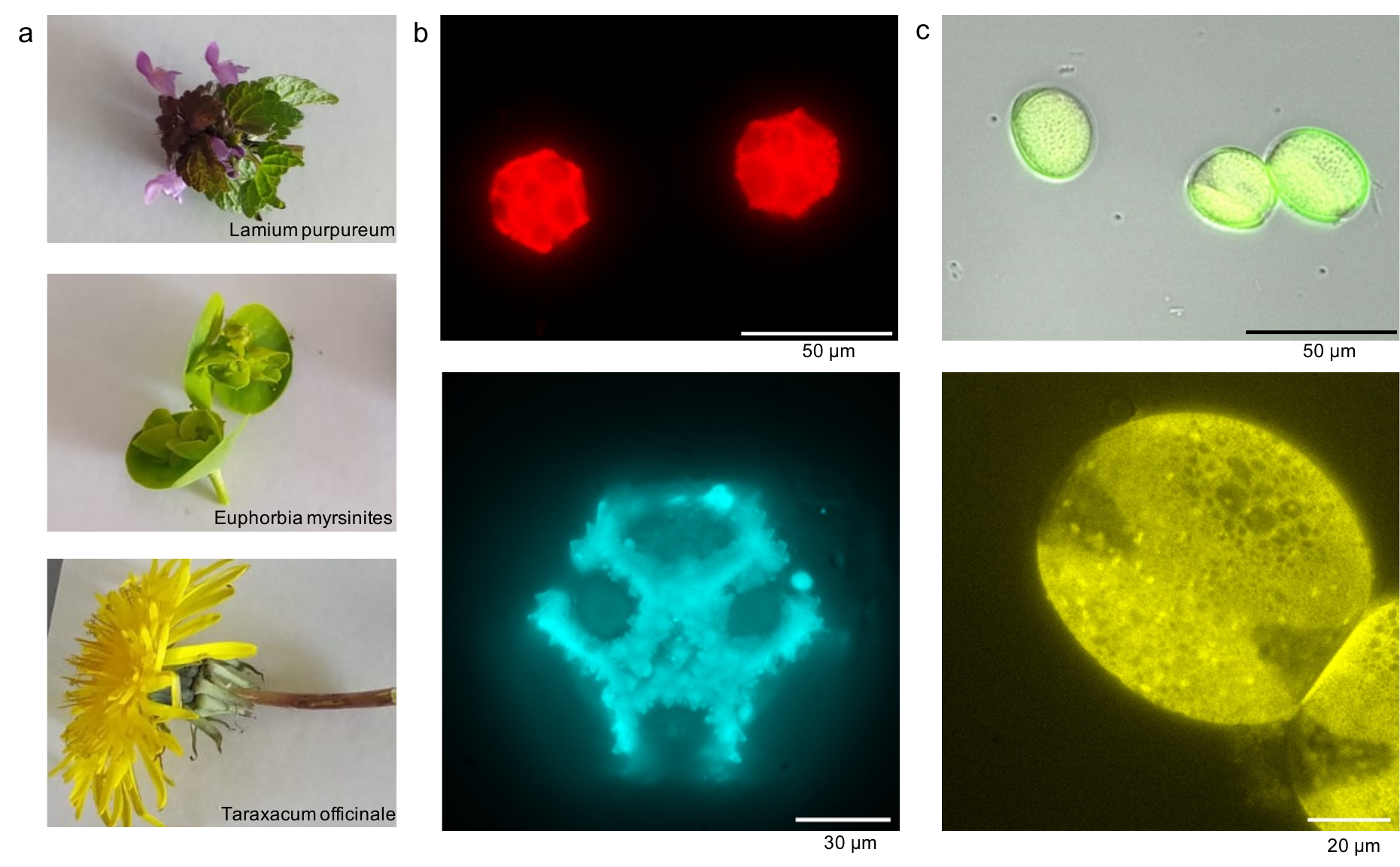}
    \caption{Examples of the plants include Taraxacum officinale, Lamium purpureum, and Euphorbia myrsinites, as well as isolated cells subjected to Flat Cell Imaging. a) Photos of the flowers prior to extraction of the cells. b,c) Examples of microscopy images the isolated cells taken without flattening and with flat cell imaging method.}
    \label{fig:enter-label}
\end{figure}
\newpage
\subsection*{Recovery of the flattened cells}
\begin{figure}[h!]
    \centering
    \includegraphics[width=\textwidth]{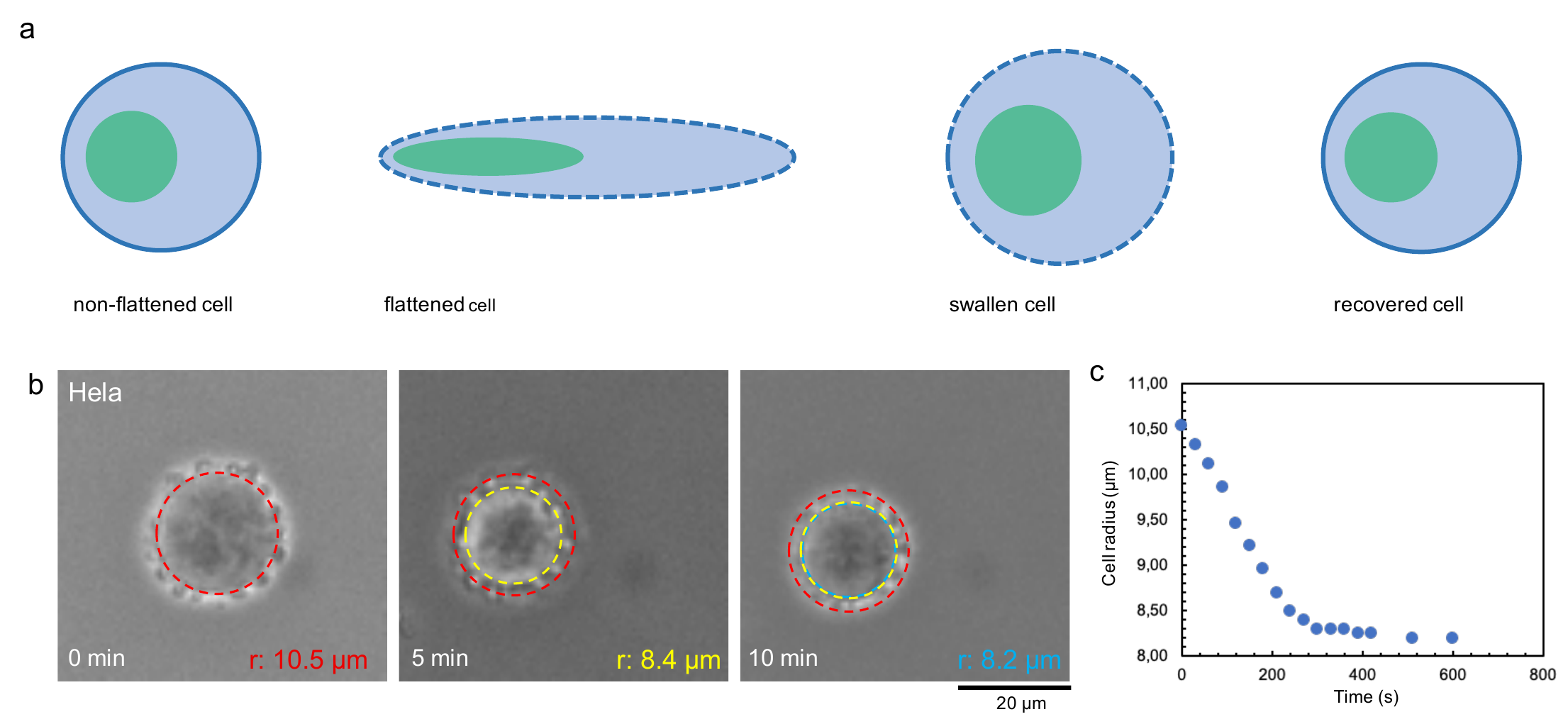}
    \caption{Recovery of the Hela cells after being flattened for 15 mins. a) Schematics of the cell cell flattening and recovery, the flattening results in a quasi-2D shape of the cell and potentially the cell membrane disruption which is shown as discontinuous line. Upon collection the cells the retain their 3D shape and recover to the original size. b) Tracking of a single Hela cell after released from the 2D flattening which shows size reduction over the course of time. Dash-lined circles represent the periphery of the cell. c) Size measurements (radius r) of the cell under recovery phase. most of the recovery happens around 5 mins of time.}
    \label{fig:enter-label}
\end{figure}
\newpage
\subsection*{Life cycle study of flattened \textit{Dictyostelium discoideum (D.d)} cells}
\begin{figure}[h!]
    \centering
    \includegraphics[width=\textwidth]{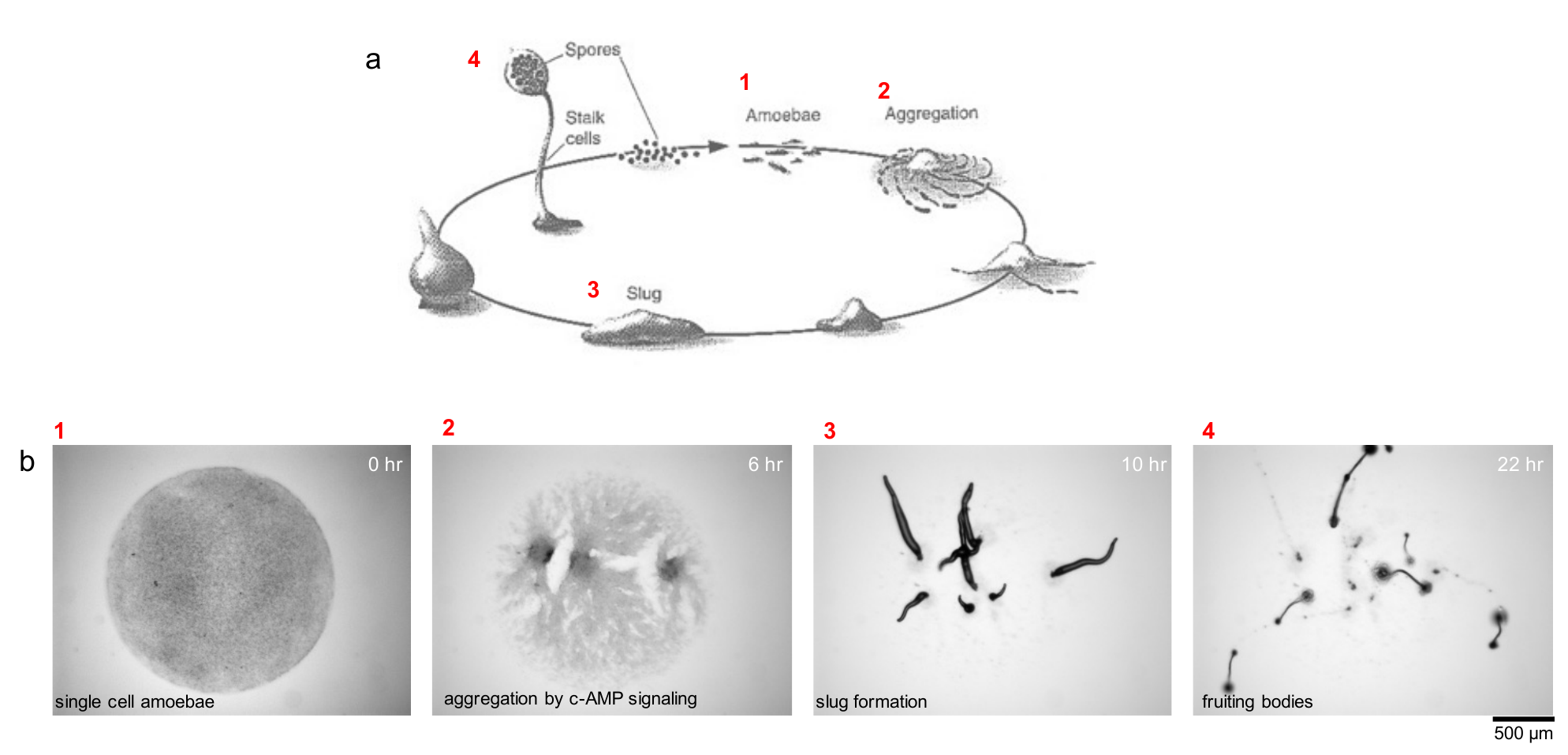}
    \caption{Visualization of the life cycle of \textit{Dictyostelium Discodium} (\textit{D.d}) after being flattened for 15 mins and subjected to starvation. a) Schematics of the life cycle of social amoeba, \textit{D.d}, with highlighted single cell, aggregation, slug and fruiting body stages, marked by the numbers (1-4). b) Phase contrast microscopy images the 4 steps in the life cycle, (1) single amoebae cells (the circle shows a droplet of the cell on agar), (2) aggregation phase, (3) \textit{D.d} slug formation, and (4) fruiting body, respectively.}
    \label{fig:enter-label}
\end{figure}
\newpage
\subsection*{Cell differentiation data}
\begin{figure}[h!]
    \centering
    \includegraphics[width=\textwidth]{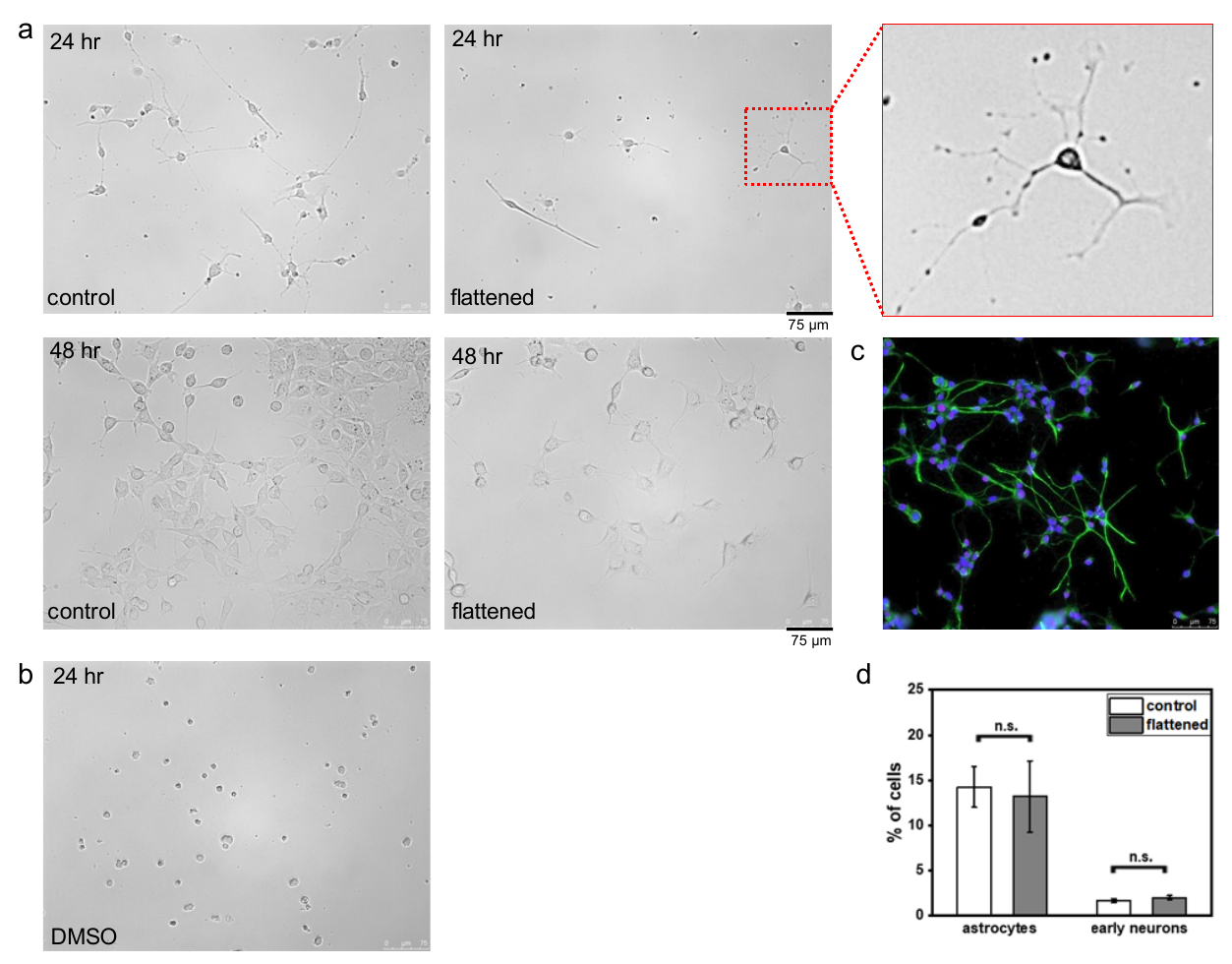}
    \caption{NSCs cell differentiation study of the flattened cells subjected to Flat Cell Imaging. a) compares the differentiation of the NSC as control and flattened after 24 and 48 hours of re-culturing (the inset view highlights the differentiation of the NSCs to neurons). Both of the control and flattened cells were subjected to EVs to start the differentiation. b) represents the image of the control cells subjected to DMSO which result in cell death, it is shown as a reference image to highlight the shape in a dead form. c) }
    \label{fig:enter-label}
\end{figure}
\newpage
\subsection*{Proliferation of the flattened cells}
\begin{figure}[h!]
    \centering
    \includegraphics[width=0.85\textwidth]{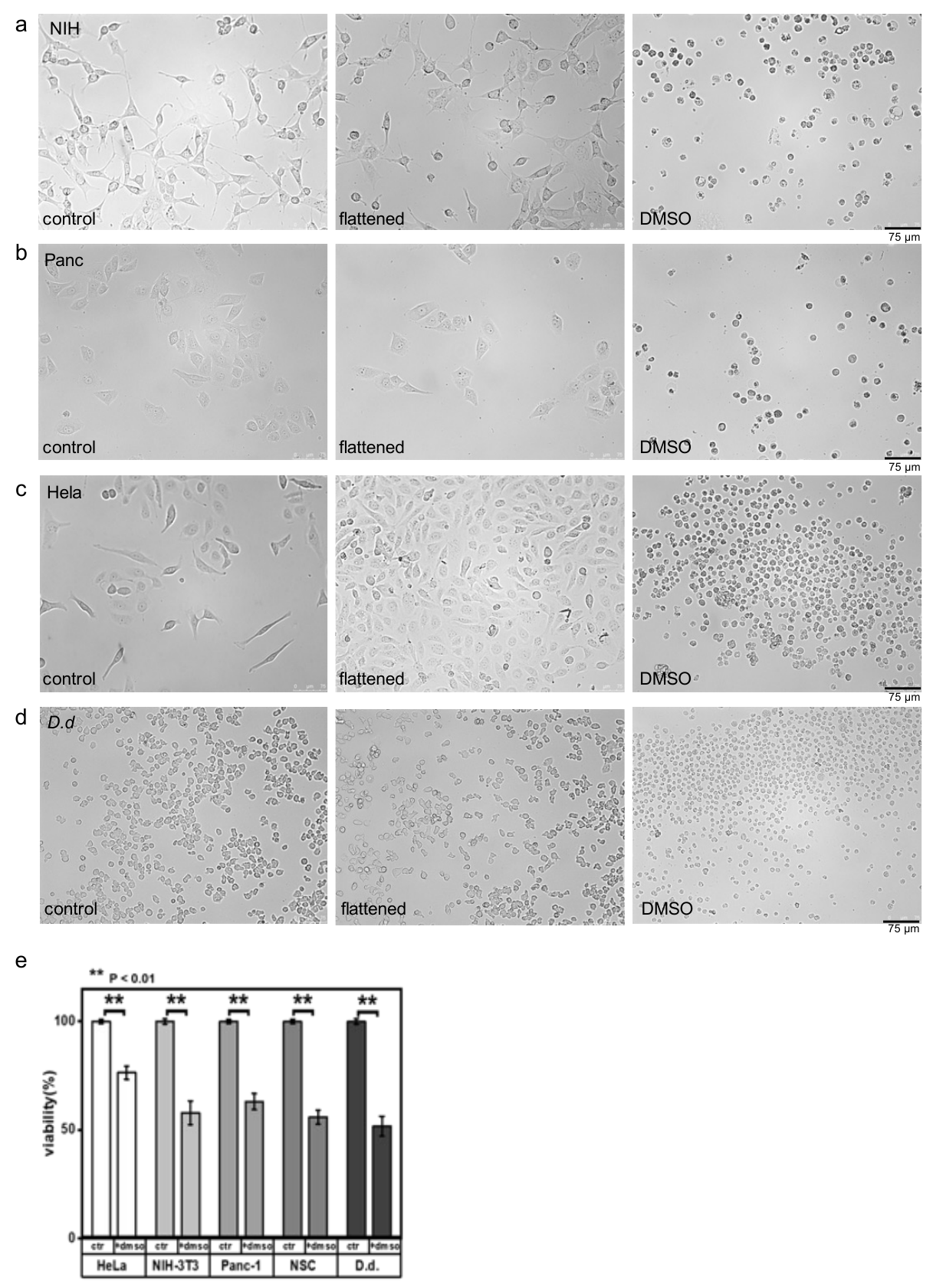}
    \caption{Proliferation of the flattened cells for various mammalian and amoeba cells. a-d) represents a comparison of cells as control (non-flattened cells), cells being re-cultured after being flattened and cells treated with Dimethyl sulfoxide (DMSO). DMSO treatment is used to demonstrate the behavior of the dead cells which results in killing the cells. e) the graph plots the viability data for the cells after 10 mins being treated with DMSO. The cells loose their viability and proliferation capacity due to DMSO effect.}
    \label{fig:enter-label}
\end{figure}
\end{document}